\documentclass[12 pt, amsfonts, amssymb,color]{article}

\usepackage{amsmath,amssymb,amsfonts,latexsym,float,graphics,epsfig}

\begin{document}

\renewcommand\theequation{\arabic{section}.\arabic{equation}}
\catcode`@=11 \@addtoreset{equation}{section}
\newtheorem{axiom}{Definition}[section]
\newtheorem{theorem}{Theorem}[section]
\newtheorem{axiom2}{Example}[section]
\newtheorem{lem}{Lemma}[section]
\newtheorem{prop}{Proposition}[section]
\newtheorem{cor}{Corollary}[section]
\newtheorem{claim}{Claim}[section]
\newcommand{\be}{\begin{equation}}
\newcommand{\ee}{\end{equation}}

\newcommand{\equal}{\!\!\!&=&\!\!\!}
\newcommand{\rd}{\partial}
\newcommand{\g}{\hat {\cal G}}
\newcommand{\bo}{\bigodot}
\newcommand{\res}{\mathop{\mbox{\rm res}}}
\newcommand{\diag}{\mathop{\mbox{\rm diag}}}
\newcommand{\Tr}{\mathop{\mbox{\rm Tr}}}
\newcommand{\const}{\mbox{\rm const.}\;}
\newcommand{\cA}{{\cal A}}
\newcommand{\bA}{{\bf A}}
\newcommand{\Abar}{{\bar{A}}}
\newcommand{\cAbar}{{\bar{\cA}}}
\newcommand{\bAbar}{{\bar{\bA}}}
\newcommand{\cB}{{\cal B}}
\newcommand{\bB}{{\bf B}}
\newcommand{\Bbar}{{\bar{B}}}
\newcommand{\cBbar}{{\bar{\cB}}}
\newcommand{\bBbar}{{\bar{\bB}}}
\newcommand{\bC}{{\bf C}}
\newcommand{\cbar}{{\bar{c}}}
\newcommand{\Cbar}{{\bar{C}}}
\newcommand{\Hbar}{{\bar{H}}}
\newcommand{\cL}{{\cal L}}
\newcommand{\bL}{{\bf L}}
\newcommand{\Lbar}{{\bar{L}}}
\newcommand{\cLbar}{{\bar{\cL}}}
\newcommand{\bLbar}{{\bar{\bL}}}
\newcommand{\cM}{{\cal M}}
\newcommand{\bM}{{\bf M}}
\newcommand{\Mbar}{{\bar{M}}}
\newcommand{\cMbar}{{\bar{\cM}}}
\newcommand{\bMbar}{{\bar{\bM}}}
\newcommand{\cP}{{\cal P}}
\newcommand{\cQ}{{\cal Q}}
\newcommand{\bU}{{\bf U}}
\newcommand{\bR}{{\bf R}}
\newcommand{\cW}{{\cal W}}
\newcommand{\bW}{{\bf W}}
\newcommand{\bZ}{{\bf Z}}
\newcommand{\Wbar}{{\bar{W}}}
\newcommand{\Xbar}{{\bar{X}}}
\newcommand{\cWbar}{{\bar{\cW}}}
\newcommand{\bWbar}{{\bar{\bW}}}
\newcommand{\abar}{{\bar{a}}}
\newcommand{\nbar}{{\bar{n}}}
\newcommand{\pbar}{{\bar{p}}}
\newcommand{\tbar}{{\bar{t}}}
\newcommand{\ubar}{{\bar{u}}}
\newcommand{\utilde}{\tilde{u}}
\newcommand{\vbar}{{\bar{v}}}
\newcommand{\wbar}{{\bar{w}}}
\newcommand{\phibar}{{\bar{\phi}}}
\newcommand{\Psibar}{{\bar{\Psi}}}
\newcommand{\bLambda}{{\bf \Lambda}}
\newcommand{\bDelta}{{\bf \Delta}}
\newcommand{\p}{\partial}
\newcommand{\om}{{\Omega \cal G}}
\newcommand{\ID}{{\mathbb{D}}}
\newcommand{\pr}{{\prime}}
\newcommand{\prr}{{\prime\prime}}
\newcommand{\prrr}{{\prime\prime\prime}}
\title{On the quest for generalized Hamiltonian descriptions of $3D$-flows generated by curl of a vector potential}

\author{O\u{g}ul Esen\footnote{E-mail: oesen@gtu.edu.tr}\\
Department of Mathematics\\ Gebze Technical University, \\ 41400, Gebze, Kocaeli, Turkey \\ \\ Partha Guha\footnote{E-mail: partha@bose.res.in}\\
SN Bose National Centre for Basic Sciences \\
JD Block, Sector III, Salt Lake \\ Kolkata 700098,  India \\
}

\date{ }

 \maketitle

\smallskip

\smallskip

\begin{abstract}
\textit{We study Hamiltonian analysis of three-dimensional advection flow $\mathbf{\dot{x}}=\mathbf{v}({\bf x})$
of incompressible nature $\nabla \cdot {\bf v} ={\bf 0}$ assuming that
dynamics is generated by the curl of 
a vector  potential $\mathbf{v} = \nabla \times \mathbf{A}$. More concretely, we elaborate Nambu-Hamiltonian and bi-Hamiltonian characters of such systems under the light of vanishing or non-vanishing of the quantity $\mathbf{A} \cdot \nabla \times \mathbf{A}$. We present an example (satisfying  $\mathbf{A} \cdot \nabla \times \mathbf{A} \neq 0$) which can be written as in the form of Nambu-Hamiltonian and bi-Hamiltonian formulations. We present another example (satisfying  $\mathbf{A} \cdot \nabla \times \mathbf{A} = 0$) which we cannot able to write it in the form of a Nambu-Hamiltonian or bi-Hamiltonian system. On the hand, this second example can be manifested in terms of  Hamiltonian one-form and
yields generalized or vector 
Hamiltonian equations $\dot{x}_i = - \epsilon_{ijk}{\partial \eta_j}/{\partial x_k}$.}
\end{abstract}

\smallskip


\smallskip


\smallskip


\section{Introduction}

Hamiltonian analysis of finite dimensional systems has a huge literature and still attracts deep attention of many autors.  
There are several types of 3D flows which you can interpret in terms of Nambu-Poisson Hamiltonian dynamics. As it is well known, the generalized force field can be decomposed into the sum of two vector fields. One  which is equal to minus of
the gradient of some potential function whereas the other does not
come from a potential. This decomposition resembles the
decomposition of a vector field into a curl-free (irrotational)
component and a solenoidal (divergence-free) component
arising from the celebrated Helmholtz theorem. A particular instance of this Helmholtz-like decomposition is the one where the irrotational and the divergence-free parts are orthogoal \cite{Tsallis}. 

We start with a three dimensional system in form
\begin{equation}\label{one}
\mathbf{\dot{x}}=\mathbf{v}({\bf x},t),
\end{equation}
where we denote a three-tuple by $\mathbf{x}=(x,y,z)$, describing the evolution with time $t$ of the spatial 
position 
${\bf x}$ of a fluid particle under the action of the fluid velocity field $\mathbf{v}({\bf x},t)$, known as
the advection equation, which can be described briefly as follows \cite{TangBoozer, ThiffeaultBoozer}. 
The transport of a passive scalar $\psi$ embedded in a fluid flow governed by the
ideal advection equation
$$ \frac{\partial \psi}{\partial t} + {\bf v}({\bf x},t) = 0 \qquad \Leftrightarrow \qquad  
\frac{d\psi}{dt} = 0.
$$
Since the passive scalar is frozen
into the fluid element, the distribution function $\psi$ at arbitrary time can be found to be
$$\frac{d{\bf x}(\xi,t)}{dt} = {\bf v}({\bf x},t)$$ with the initial condition ${\bf x}(\xi, t=0) = \xi$. In general, (\ref{one}) is a nonlinear dynamical system capable of exhibiting chaotic dynamics,
then this flow is a mixing flow over some region of Lagrangian topology. The flow is mixing means 
whether the flow trajectory is globally ergodic, i.e., trajectory visits every point in a closed domain.

If the flow is incompressible $\nabla \cdot {\bf v} = 0$, then the dynamical system (\ref{one}) is conservative.
The representation of divergence-free vector fields as curls in two and three dimensions has been studied in
\cite{Barbarosie}. In such a situation ${\bf v}$ can be represented as ${\bf v}({\bf x}) = \nabla \gamma_1 \times
\nabla \gamma_2$, which can be manifested in terms of Nambu-Hamiltonian form \cite{CGR,Morando}.
In this paper we show an example involving an additional condition on the dynamics of the flow 
${\bf v}({\bf x}) = \nabla \times {\bf A}$,
i.e., null helicity condition etc, then Hamiltonian formalism can not be given in terms of Nambu-Hamiltonian.
This can be restored if we break the symmetry of the null helicity. Accordingly, we notice one can grossly divide these into two categories.

In the first category, there are 3D flows satisfying both the divergence free condition and 
the integrability condition, that is
\begin{equation} 
\nabla\cdot\mathbf{v}=0,\qquad \mathbf{v} \cdot \nabla\times \mathbf{v} = \mathbf{0}.
\end{equation} 
Here, the integrability condition $\mathbf{v} \cdot \nabla\times \mathbf{v} = 0$ of the vector field $\mathbf{v}$ 
is defined up to a multiplier $\mu$. This can be shown as follows.
Let us consider a system of equation 
$$ \dot{x} = P(x,y,z), \qquad \dot{y} = Q(x,y,z),\qquad  \dot{z} = R(x,y,z). $$
Then a direct calculation follows
\begin{eqnarray*}
&&\mu \mathbf{v} \cdot \big( \nabla \times \mu \mathbf{v}) \\ &&\qquad = \mu P\big( \frac{\partial}{\partial y}(\mu R) - 
\frac{\partial}{\partial z}(\mu Q) \big) + \mu Q\big( \frac{\partial}{\partial z}(\mu P) - 
\frac{\partial}{\partial x}(\mu R) \big) + \mu R\big( \frac{\partial}{\partial x}(\mu Q) - 
\frac{\partial}{\partial y}(\mu P) \big)
\\&&\qquad 
= \mu^2 \Big( P \big(\frac{\partial R}{\partial y} - \frac{\partial Q}{\partial z}\big) + 
Q \big(\frac{\partial P}{\partial z} - \frac{\partial R}{\partial x}\big) + R \big(\frac{\partial Q}{\partial x} - 
\frac{\partial P}{\partial y}\big) \Big) \\&&\qquad  = \mu^2 ( \mathbf{v} \cdot \nabla\times \mathbf{v} ) = 0.
\end{eqnarray*}
Note that, in the second line the other terms are identically vanishing. It is immediate to see that, if $\nabla\cdot\mathbf{v}=0$ we have that
\begin{equation} 
\mathbf{v} = K\nabla H, \qquad \text{or} \qquad \nabla \times \mathbf{v} = \nabla K \times \nabla H,
\end{equation}
for two real valued functions $K$ and $H$. As an example for this case, consider the Euler top equation
\begin{equation} \label{Euler-Eq}
\dot{x} = Ayz, \qquad \dot{y} = Bxz, \qquad \dot{z} = Cxy,
\end{equation} 
where $A$, $B$ and $C$ being constants. The divergence free condition generates Liouville’s theorem of Nambu mechanics, 
therefore the state space can be regarded as an incompressible fluid. 
It is interesting to notice that if the coefficients $A$, $B$ and $C$ in the system (\ref{Euler-Eq}) are all equal to each other, then $\nabla\times \mathbf{v}$ vanishes identically. This motivates the following special case of such flows 
$$ \nabla \cdot \mathbf{v} = 0, \qquad \nabla\times \mathbf{v} = \mathbf{0}.$$
As an example, consider the Lagrange system, the case $A=B=C=1$,
\begin{equation}
\dot{x} = yz, \qquad \dot{y} = xz, \qquad \dot{z} = xy,
\end{equation}
and 3D circle map equation $\mathbf{\dot{x}} = x^2$. These flows are the usual (conservative) potential flows  $\mathbf{v} = - \nabla V$ generating by a potential function $V$.  

As an another category, let the generator vector field $\mathbf{v}$ in a $3D$ system is curl of a potential field $\mathbf{A}$. In general, the  integrability condition is not fulfilled. In this case, we have
\be \label{intv}
\mathbf{v}=\nabla \times \mathbf{A},\qquad \nabla\times \mathbf{v}\neq \mathbf{0},
\ee
where $\mathbf{A}$ is the potential vector field. Motivating from the Hamiltonian curl forces in two dimensions discussed in \cite{BS466, BS2,BS3}, our focus in this study is to elaborate Hamiltonian analysis of three dimensional differential system satisfying \eqref{intv}. 
In this present work, we distinguish two subcategories according to the satisfaction of the potential vector field $\mathbf{A}$ of the integrability condition. 

In the first subcategory, we study $$ \mathbf{A} \cdot \nabla \times \mathbf{A} \neq 0.$$ 
It is known from the standard text book that for a 
solenoidal and continuously differentiable vector function such as a 
vector field $\mathbf{v}$ can be expressed as a cross-product of two gradients, $$\mathbf{v} = \nabla H \times \nabla K,$$
where $H$ and $K$ are assumed to smooth functions \cite{Barbarosie}. The standard proof assumes that there exist two families of surfaces, 
$H(x, y, z) = c_1$ and $K(x, y, z) = c_2$, for which all the flow lines are confined to the surface of intersections.
This is the essential ingredient of the Nambu mechanics too. It is worth noting that
if we start with $ \mathbf{A} = H\nabla K$ or $\mathbf{v} = \nabla H \times \nabla K$, the scalar product
$\mathbf{A}\cdot \mathbf{v} = 0$. It is known that the
vector potential corresponding to vector field $\mathbf{v}$ can be presented in the Clebsch representation \cite{Biskamp}
form 
\be
\mathbf{A} = - K \nabla H + \nabla M,
\ee
where the function $M$ must be multi-valued. This implies that the function
$M$ has a surface $\Sigma$ inside the volume $\Omega$ where it has a jump, then the contribution from the jump
surface $\Sigma$ is added to the integral over $\partial \Omega$, which results in the nonzero helicity.
It is a challenging question to find the function $M$ and to understand why it has to be multi-valued \cite{Biskamp,SKB}. 
In three dimensions, the geometric aspects can be discussed in terms of bi-Hamiltonian and Nambu-Hamiltonian formulations apart from the two dimensional ones. Accordingly, we shall discuss the Hamiltonian characters of the systems (\ref{intv}) in terms of the bi-Hamiltonian and the Nambu-Hamiltonian frameworks. 
We give two different examples for this subcategory, one possesses the Nambu structure whereas the other one has
the almost Nambu structure involving a multiplier.

The second subcategory covers a divergence free flow
satisfying integrability condition $$ \mathbf{A} \cdot \nabla \times \mathbf{A} = 0.$$ We should expect if these 
conditions are satisfied then $\mathbf{A} = H\nabla K$, then it would yield  Nambu Hamiltonian system, unfortunately
we fail to get it. The closed two form $\omega$ associated to dynamics is the product of two one-forms 
$\omega = J_1 \wedge J_2$, unlike the previous cases in this $J_1$ and $J_2$ are not exact.
 The Euler potentials
could be discontinuous, although the vector potential and $\mathbf{v} = \nabla \times \mathbf{A}$ 
might not have discontinuities on the intersection of surfaces.
In this paper, we demonstrate if we deform
the system to another which satisfies divergence free condition and  $\mathbf{A}  \cdot \nabla \times \mathbf{A} \neq 0$,
then it becomes Nambu-Hamiltonian.

Our motivation to study such flows coming
from the work of Berry and Shukla on curl forces we have studied in this paper
curl velocities on 3D spaces. We have studied 3D vector fields generated by a curl of a vector
potential. In general, the theory for 3D systems is less well-developed \cite{Wiggins}, there are many unanswered
questions regarding the nature of dynamics in 3D flows. The flows which have been investigated in this
paper are also divergence free vector fields
like Berry-Shukla, and these could be either Hamiltonian or non-Hamiltonian.

In this point, we want to make a final remark that 3D flows possess  extremely diverse characters. Consider, for example, the following 3D flow, called as  
SIR equation, given by $$\dot{S} = -rSI, \qquad \dot{I} = rSI - aI, \qquad \dot{R} = aI,$$ which neither satisfies
$\nabla \cdot \mathbf{v} = 0$ nor $\mathbf{v}=\nabla \times \mathbf{A}$, but still one can express
it in terms of Nambu Hamiltonian form.

\bigskip 

\noindent \textbf{Organization of the paper.} In Section 2, we present  basics of Hamiltonian realizations of $3D$ systems. In Section 3 we exhibit some illustrations that posses Nambu-Hamiltonian and bi-Hamiltonian character. In Section 4, we examine a counter example which cannot in the Nambu-Hamiltonian and bi-Hamiltonian formalism.
 
\section{Hamiltonian Analysis of Three Dimensional Velocities}

\subsection{Three Dimensional Hamiltonian Systems}

Let $(\mathcal{P},\{\bullet,\bullet\})$ be a $3$-dimensional Poisson manifold equipped with Poisson bracket $\{\bullet,\bullet\}$ satisfying the Jacobi identity. The Hamilton's equation generated by a Hamiltonian function is 
\begin{equation}
\mathbf{\dot{x}}=\{\mathbf{x},H\},
\end{equation}
for local coordinates $(\mathbf{x})$ on $\mathcal{P}$. 
In $3$-dimensions, we can replace the role of a Poisson bracket with a Poisson vector $\mathbf{J}$, \cite{EsGhGu16, Gum1, GuNu93}. In this case, the Jacobi identity turns out to be the following  equation
\begin{equation} \label{jcbv}
\mathbf{J}\cdot(\nabla\times\mathbf{J})=0,
\end{equation}
whereas the Hamilton's equation takes
the particular form 
\begin{equation} \label{HamEq3}
\mathbf{\dot{x}}=\mathbf{J}\times\nabla H.
\end{equation}
Here, $H$ is a Hamiltonian function defined on $\mathcal{P}$, and $\nabla H$ is the gradient of $H$.
The following theorem is exhibiting all possible solutions of Jacobi identity given in (\ref{jcbv}) so that characterizes all Poisson Poisson structures in $3$-dimensions, \cite{AGZ,HB1,HB2,HB3}.
\begin{theorem} \label{ss}
The general solution of the vector equation (\ref{jcbv}) is 
$\mathbf{J}= \left({1}/{M} \right) \nabla F$ for arbitrary functions $M$
and $F$. \end{theorem}
The existence of scalar multiple ${1}/{M}$ in the solution is a manifestation of  conformal invariance of the identity (\ref{jcbv}). In the literature, $M$ is called as the Jacobi's last multiplier \cite{Jac1,Jac2}. In this picture, a Hamiltonian system has the following generic form
\begin{equation} \label{HamEq4}
\mathbf{\dot{x}}= \frac{1}{M} \nabla F \times \nabla H.
\end{equation}

A dynamical system is bi-Hamiltonian if it admits two different Hamiltonian
structures
\begin{equation}
\mathbf{\dot{x}}= \{\mathbf{x},H_2\}_{1}=\{\mathbf{x},H_1\}_{2},\label{biHam}%
\end{equation}
with the requirement that the Poisson brackets $\{\bullet,\bullet \}_{1}$ and $\{\bullet,\bullet \}_{2}$ be
compatible \cite{Fe94,OLV}.
Recalling the system (\ref{HamEq4}), we arrive at that a Hamiltonian system in the form (\ref{HamEq4}) is bi-Hamiltonian
\begin{equation} \label{bi-Ham}
\mathbf{\dot{x}}= \frac{1}{M} \nabla F\times\nabla H=\mathbf{J}_{1}\times\nabla H=\mathbf{J}_{2}\times\nabla
F,
\end{equation}
where, the first Poisson vector $\mathbf{J}_{1}$ is given by $(1/M)\nabla F$ whereas the second Poisson vector $\mathbf{J}_{2}$ is given by $-(1/M)\nabla H$. The following theorem determine the Hamiltonian picture of three dimensional dynamical systems admitting an integral invariant. For the proof, we refer \cite{EsGhGu16,Gao}.
\begin{theorem} \label{ss2}
A three dimensional dynamical system $\mathbf{\dot{x}}=\mathbf{v}(\bf{x})$ having a
time independent first integral is bi-Hamiltonian if and only if there exist
a Jacobi's last multiplier $M$ which makes $M\mathbf{v}$ divergence free.
\end{theorem}

\subsection{$3D$ Nambu-Poisson manifolds}

Let $\mathcal{P}$ be a three dimensional manifold. A Nambu-Poisson bracket of order $3$ is a ternary operation, denoted by $\{\bullet ,\bullet ,\bullet
\}$, on the space of smooth functions, satisfying both the generalized
Leibniz identity
\begin{equation} \label{GLI}
\left\{ F_{1},F_{2},FH\right\} =\left\{ F_{1},F_{2},F\right\} H+F\left\{
F_{1},F_{2},H\right\}
\end{equation}%
and the fundamental (or Takhtajan) identity
\begin{equation}
\left\{ F_{1},F_{2},\{H_{1},H_{2},H_{3}\}\right\}
=\sum_{k=1}^{3}\{H_{1},...,H_{k-1},\{F_{1},F_{2},H_{k}\},H_{k+1},...,H_{3}\},
\label{FI}
\end{equation}%
for arbitrary functions $F,F_{1},F_{2},H,H_{1},H_{2}$, see \cite{Nambu, Ta}. 

Assume that $(\mathcal{P},\{\bullet ,\bullet ,\bullet
\})$ be a Nambu-Poisson manifold. For a pair $(H_1,H_2)$ of Hamiltonian functions, the associated Nambu-Hamiltonian vector field $X_{H_1,H_2}$ is defined through 
\begin{equation}\label{NambuHamVF}
  X_{H_1,H_2}(F)=\{F,H_1,H_2\}.
\end{equation}
The distribution of the Nambu-Hamiltonian vector fields are in involution and defines a foliation of the manifold $\mathcal{P}$. A
dynamical system is called Nambu-Hamiltonian with a pair $(H_1,H_2)$ of Hamiltonian
functions if it can be recasted as%
\begin{equation}
\mathbf{\dot{x}}=\left\{\mathbf{x} ,H_{1},H_{2}\right\} .  \label{NHamEqn}
\end{equation}

If a system is in the Nambu-Hamiltonian form (\ref{NHamEqn}) then by fixing one of the Hamiltonian functions in the pair $(H_1,H_2)$, we can write it in the bi-Hamiltonian form as well
\begin{equation} 
\mathbf{\dot{x}}=\left\{ \mathbf{x},H_{1}\right\} ^{H_{2}}=\left\{ \mathbf{x}
,H_{2}\right\} ^{H_{1}}  \label{NH-2Ham}
\end{equation}%
where the brackets $\{\bullet ,\bullet \}^{H_{2}}$ and $\{\bullet ,\bullet
\}^{H_{1}}$ are compatible Poisson structures defined by
\begin{equation}
\left\{ F,H\right\} ^{{H}_{2}}=\left\{ F,H,H_{2}\right\} \text{, \ \ \ }%
\left\{ F,H\right\} ^{{H}_{1}}=\left\{ F,H_{1},H\right\} ,  \label{Pois}
\end{equation}%
respectively.

Let $\mathcal{P}$ be a 3 dimensional manifold equipped with a non-vanishing volume manifold $\mu$. Then the following identity 
\begin{equation}\label{NP-Vol} 
  \{F_1,F_2,F_3\}\mu=dF_1\wedge dF_2 \wedge dF_3
\end{equation}
defines a Nambu-Poisson bracket on $\mathcal{P}$ \cite{Ga96,Gu01}. In this case, the equation (\ref
{NambuHamVF}) relating a Hamiltonian pair $(H_1,H_2)$ and a Nambu-Hamiltonian vector field $X_{H_1,H_2}$ can be written, in a covariant formulation, as 
\begin{equation} \label{NHcf}
\iota_{X_{H_1,H_2}}\mu = dH_1 \wedge dH_2,
\end{equation}
where $\iota$ is the interior derivative. 
We call (\ref{NHcf}) as the Nambu-Hamilton's equations \cite{Fe92}.
Note that, by taking the exterior derivative of both hand side of (\ref{NHcf}), we arrive at the preservation of the volume form by the Nambu-Hamiltonian vector field, that is
\begin{equation} \label{NH-vf-mu}
\mathcal{L}_{X_{H_1,H_2}}\mu =0.
\end{equation}
Integration of this conservation law gives that the flow of a Nambu-Hamiltonian vector field is a volume preserving diffeomorphism.

Consider a local frame (called as the standard basis) given by a three-tuple $(u,v,w)$ such that the volume form is 
\begin{equation}\label{volume}
 \mu= du \wedge dv \wedge dw.
\end{equation}
In this picture the Nambu-Poisson three-vector takes the particular form
\begin{equation}\label{Nambu}
N= \frac{\partial}{\partial u} \wedge  \frac{\partial}{\partial v}  \wedge  \frac{\partial}{\partial w}.
\end{equation}
Locally, the Nambu-Hamiltonian vector field $X_{H_1,H_2}$ defined in (\ref{NHcf}) for a pair $(H_1,H_2) $ of Hamiltonian functions can be computed as
\begin{equation} \label{HbhVF}
X_{H_1,H_2} =  \{H_1,H_2\}_{u,v} {\frac{%
\partial}{\partial w}} + \{H_1,H_2\}_{v,w} {\frac{\partial}{\partial u}} + 
\{H_1,H_2\}_{w,u} {\frac{\partial}{\partial v}},
\end{equation}
where the coefficients are, for example,
\begin{equation} \label{pbh}
\{H_1,H_2\}_{a,b} = {\frac{\partial H_1}{\partial a}}{\frac{\partial H_2}{%
\partial b}} - {\frac{\partial H_1}{\partial b}}{\frac{\partial H_2}{%
\partial a}}.
\end{equation}

For the particular case of a three dimensional Euclidean space, the present discussion reduces to the following form. Let $F_1$, $%
F_{2}$ and $F_{3}$ be three real valued functions, and consider the triple product
\begin{equation}
\left\{ F_{1},F_{2},F_{3}\right\} = \nabla F_1 \cdot \nabla F_{2}\times
\nabla F_{3}  \label{NambuPois}
\end{equation}%
of the gradients of these functions. It is evident that the bracket \eqref{NambuPois} is a Nambu-Poisson bracket with corresponding Nambu--Poisson three--vector field in the standard form (\ref{Nambu}). The Nambu-Hamiltonian vector field presented in (\ref{HbhVF}) takes the particular form
\begin{equation*}
X_{H_1,H_2} = \nabla H_{1}\times\nabla H_{2}.
\end{equation*}
It follows that, the Nambu-Hamilton's equations (\ref{NHamEqn}) turn out to be
\begin{equation} \label{NH3D}
\mathbf{\dot{x}}=\left\{\mathbf{x} ,H_{1},H_{2}\right\}= \nabla H_{1}\times\nabla H_{2}.%
\end{equation}
The bi-Hamiltonian character of this system can easily be observed by employing (\ref{NH-2Ham}).
The divergence of (\ref{NH3D}) generates Liouville’s theorem of Nambu mechanics:
$$ \nabla \cdot \mathbf{\dot{x}} =  \nabla \cdot \big(\nabla H_{1}\times\nabla H_{2}\big) = 0.
$$
The main advantage of using the two
conserved quantities (Hamiltonians) in Nambu formulation, is the representation of the phase space trajectory
as intersection line of two surfaces based on the conserved quantities. This geometric
application illustrates the kind of motion without explicitly solving the equations of motion, this 
is the key feature of the (maximal) superintegrability.

\subsection{Three Dimensional Systems} 

Our motivation stems from the equation of the chaotic advection of dye
\be \label{vsys}
\dot{\bf x} = {\bf v}, \qquad \nabla\times \mathbf{v}\neq\mathbf{0}
\ee
where the velocity field is assumed to have
been determined a priori and satisfies the incompressibility condition $\nabla \cdot {\bf v} = 0$. 
All the flows we consider in this section satisfy the Frobenius integrability condition. 
An handy way to describe Frobenius integrability condition has been prescribed by 
Ollagnier and Strelcyn \cite{OS}. Consider a
smooth one form $J$ corresponding to a first integral $I$ of a smooth dynamical vector 
field $\mathbf{v}$ defined on ${\Bbb R}^3$. It is tautological that the volume form $\Omega$ and
$J$ satisfy $\Omega \wedge J = 0$. If we take
interior product with respect to the vector field $\mathbf{v}$, we get $i_{\mathbf{v}}\Omega \wedge J = 0$.
We now state a general remark that a
smooth function $I$ is a first integral of a smooth vector field $\mathbf{v}$ defined on ${\Bbb R}^3$ 
if and only if $dI \wedge i_{\mathbf{v}}\Omega = 0$. The advection equations are non-integrable in general.
A large class of conservative systems, which exhibit
chaotic behavior, has a Hamiltonian representation.
Two of the well-known examples are the magnetic
field $B$ and the velocity field $v$ of a divergence free
fluid. 

Let us start by considering a vector potential $\mathbf{A}=(A_x,A_y,A_z)$, and the following system of equations
\be \label{se}  
{\bf \dot x} = \mathbf{v}(x)= \nabla \times \mathbf{A}(x).
\ee
It is easy to observe now that the vector potential $\mathbf{A}$ is far from being unique since 
\be \nabla\times (\mathbf{A} + \nabla \phi) = \nabla\times \mathbf{A} + \nabla\times \nabla \phi =  \nabla\times \mathbf{A}  \ee
for an arbitrary real valued function $\phi$. This is the gauge invariance of the dynamics.
It is immediate to observe that $\nabla \times \mathbf{v}$ of the velocity field does not necessarily zero for arbitrary vector potential $\mathbf{A}$. 

In terms of the differential forms, the picture is as follows. Consider a differential one-form
\begin{equation} \label{alp}
\alpha=\mathbf{A}(x)\cdot d\mathbf{x},
\end{equation}
so that we have
\begin{equation}
*d \alpha=\nabla \times \mathbf{A}\cdot d\mathbf{x}=\mathbf{v}(\mathbf{x})\cdot d\mathbf{x},
\end{equation} 
where $*$ is the Hodge star operator with respect to the Euclidean norm.
Alternatively, start with a volume form
$\Omega = dx \wedge dy \wedge dz$. If we contract with the dynamical vector field 
$\hat{\mathbf{v}} = v_i {\partial}/{\partial x_i}$ we obtain the two form $d\alpha$. Therefore, the equations of motion presented in (\ref{se}) can also be expressed in the following form
\be \label{v-2}
\mathbf{\dot{x}}=\ast(d\alpha \wedge  d\mathbf{x}).
\ee



In $3$-dimensions, any two-from is decomposable. Accordingly, we write the exact two-form $d\alpha$ as the wedge product of two one-forms $J_1$ and $J_2$ that is 
\be \label{dv}
d\alpha = dv_x \wedge dx + dv_y \wedge dy + dv_z \wedge dz = J_1 \wedge J_2.
\ee
Here, $J_1$ and $J_2$ are two integral invariants of the system.
If both $J_1$ and $J_2$ are closed then by Poincar\'e lemma, we arrive at two first integrals $I_1$ and $I_2$ satisfying $J_1 = dI_1$ and $J_2 = dI_2$ respectively. If this is the case, then $3$-dimensional phase flow can be described by means
of the first integrals. From geometric point of view, this gives that a solution to the system, that is an integral curve, can be realized as the intersection of two level surfaces defined by the first integrals $I_1$ and $I_2$. So that we can write the system (\ref{vsys}) as follows
\be \dot{\bf x} = \nabla I_1 \times \nabla I_2. \ee
In this case, we can write $\mathbf{A}=I_1\nabla I_2$. Note that this representation of the dynamics coincide with the standard form of the Nambu-Hamilton equations exhibited in (\ref{NH3D}). Further, by employing (\ref{NH-2Ham}), we see that this description is bi-Hamiltonian as well.

We will illustrate two types of flows; one kind of 3D flows yield Nambu-Hamiltonian mechanics and they satisfy
$\mathbf{A}\cdot \nabla \times \mathbf{A} \neq 0$, where $\mathbf{v} = \nabla \times \mathbf{A}$. 
Other type flow does not possess Hamiltonian framework and it satisfies $\mathbf{A}\cdot \nabla \times \mathbf{A}
= 0$.

\section{Illustrations}

In this section we give two examples, one is a relatively simple one and the other one is more complicated.

\subsection{A superintegrable system }

We now consider an example of 3D system \cite{Duma2} generated by the vector potential 
\be\label{C2}
\mathbf{A}= \frac{1}{4}\left(z^2 - xy^2,x^2y - 2yz,y^2 - 2xz \right). \ee
The equations of motion (\ref{se}) can be written as 
\be\label{E2}
\dot{x} = y, \qquad  \dot{y} = z, \qquad \dot{z} = xy.
\ee
One can check trivially $\mathbf{A}\cdot (\nabla \times \mathbf{A}) \neq 0$.
It is immediate to see that $\nabla \times \mathbf{v}$ does not vanish for $\mathbf{v}=(y,x,xy)$.
The two-forms in (\ref{dv}) turns out to be
$$
zdz \wedge dx + ydy \wedge dz + xy dx\wedge dy =
J_1 \wedge J_2
$$
where the integral one-forms $J_1$ and $J_2$ can be computed to be
\be
J_1 = zdx + xdz - ydy - x^2dx, \qquad J_2 = xdx - dz
\ee
respectively. It is immediate to check the invariance of the one-forms by $\iota_{\mathbf{v}} J_i = 0$. 
Note that, $J_1 = dI_1$ and $J_2 = dI_2$ are exact with the potential functions 
\be \label{I12}
I_1 = xz - \frac{y^2}{2} - \frac{x^3}{3}, \qquad I_2 = \frac{x^2}{2} - z,
\ee
See that $I_1$ and $I_1$ are smooth first integrals of the dynamics (\ref{E2}). Thus (\ref{E2}) is a maximal superintegrable
3D dynamics generated by the curl of the vector potential $\mathbf{A}$.  

Using these two first integrals, we can express equation (\ref{E2}) in the Nambu-Hamiltonian form (\ref{NH3D}) as follows
\be
\mathbf{\dot{x}} = \{\mathbf{x},I_1,I_2\}=\nabla I_1 \times \nabla I_2
\ee
where $I_1$ and $I_1$ are the ones in (\ref{I12}).
According to (\ref{NH-2Ham}), we can represent the maximal superintegrable system (\ref{E2}) in the bi-Hamiltonian formulation 
\be
 \left(\begin{array}{c}
\dot{x}\\
\dot{y}\\
\dot{z}\\
\end{array}\right) = \left(\begin{array}{ccc}
0 & -1 & 0\\
1 & 0 & x \\
0 & -x & 0\\
\end{array}\right)\left(\begin{array}{c}
\frac{\partial I_1}{\partial x}\\
\frac{\partial I_1}{\partial y}\\
\frac{\partial I_1}{\partial z}\\
\end{array}\right) = \left(\begin{array}{ccc}
0 & x & -y\\
-x & 0 & -z+x^2 \\
y & -x & 0\\
\end{array}\right)\left(\begin{array}{c}
\frac{\partial I_2}{\partial x}\\
\frac{\partial I_2}{\partial y}\\
\frac{\partial I_2}{\partial z}\\
\end{array}\right).
\ee

\subsection{Lotka-Volterra equation}

The generalized Lotka-Volterra equation is given by
\be
\dot{x_1} = x_1(a_3x_2 + x_3 + l_1), \,\,\,\,  \dot{x_2} = x_2(a_1x_3 + x_1 + l_2) \,\,\,\,  
\dot{x_3} = x_3(a_2x_1 + x_2 + l_3).
\ee
The divergence free condition ${\partial v_i}/{\partial x_i} = 0$ imposes conditions on $a_i$
and $l_i$, such that $a_i = -1$ and $l_1 + l_2 + l_3 = 0$. Then the reduced set of equations becomes
\be
\dot{x_1} = x_1(-x_2 + x_3 + l_1), \qquad  \dot{x_2} = x_2(-x_3 + x_1 + l_2), \qquad   
\dot{x_3} = x_3(-x_1 + x_2 + l_3).
\ee
One can check directly that $ \nabla \times \mathbf{v} \neq 0$. Recast this in the vector
potential form $\mathbf{A}$ with
$$ A_i = x_1x_2x_3 + \tilde{l_i}, 
$$
 where the constants are satisfying 
 $$l_1 = \tilde{l_3} - \tilde{l_2},
 \,\,\,\, l_2 = \tilde{l_1} - \tilde{l_3} \,\,\,\, l_3 = \tilde{l_2} - \tilde{l_1}.$$
One can also check that $\mathbf{A}\cdot (\nabla \times \mathbf{A}) \neq 0$.
Just like the previous example we can express $d\ast \alpha$ in terms of $J_1$ and $J_2$ using a
multiplier $m = x_1x_2x_3$:
\be
 x_1(-x_2 + x_3 + l_1)dx_2 \wedge dx_3 + x_2(-x_3 + x_1 + l_2)dx_3 \wedge dx_1 + x_3(a_2x_1 + x_2 + l_3)dx_1 \wedge
dx_2 = m (J_1 \wedge J_2),
\ee
where 
$$ J_1 = \frac{dx_1}{x_1} +  \frac{dx_2}{x_2} +  \frac{dx_3}{x_3}, \qquad 
J_2 = dx_1 + dx_2 + dx_3 +  \frac{l_3dx_2}{x_2} - \frac{l_2dx_3}{x_3}. $$
This immediately shows $dJ_1 = dJ_2 = 0$ and the two Hamiltonians are
\be H_1 = \ln x_1 + \ln x_2 + \ln x_3, \qquad H_2 = x_1 + x_2 + x_3 + l_3\ln x_2 - l_2 \ln x_3. \ee 
Thus equations can be obtained from the standard Nambu-Hamiltonian formalism.

\section{Generalized Hamiltonian case and vector equations}

Consider the following vector potential, see for example \cite{Duma1},
\be\label{C1}
\mathbf{A} = \frac{1}{2}\left(y^3 - x^2z, z^3 - xy^2,x^3 - yz^2\right). \ee
The dynamics generated by the curl of $\mathbf{A}$ is computed to be
\be\label{E1}
\dot{x} = z^2, \qquad  \dot{y} = x^2, \qquad \dot{z} = y^2.
\ee
See that $\nabla \times \mathbf{v}$ is not vanishing for $\mathbf{v}=(z^2,x^2,y^2)$. The vector potential satisfies
$$ \mathbf{A}\cdot\nabla \times \mathbf{A}  = z^2(y^3 - x^2z) + x^2(z^3 - xy^2) + y^2(x^3 - yz^2) = 0. $$ 
Thus the helicity of the local flow is identically zero, due to this identity Sposito's result shows
that the flow streamlines are confined to flat 2D manifolds. Hence we can not get the Nambu-Hamiltonian
structure here.

\smallskip

The identity (\ref{dv}) becomes
\be\label{eqntwoform}
y^2dx\wedge dy + x^2 dz\wedge dx +z^2 dy \wedge dz = J_1 \wedge J_2,
\ee
where 
\be \label{J12} J_1 = (z^2 dy - x^2 dx), \qquad J_2 = (dz - \frac{y^2}{z^2}dx). \ee
It is easy to check that $J_1$ and $J_2$ are invariant one-forms that is  
$\iota_{\mathbf{v}}J_i = 0$.
The integrals one-forms $J_1$ and $J_2$ presented in (\ref{J12}) are not closed, i.e., 
$$ dJ_1 = 2z dz \wedge dy \neq 0, \qquad dJ_2 = 2\frac{y^2}{z^3}dz \wedge dx - 2\frac{y}{z^2} dy \wedge dx \neq 0.
$$
Hence we can not express the equation  \eqref{C1} neither in the Nambu-Hamiltonian nor the bi-Hamiltonian 
forms.

\smallskip

We can find two more such pairs like $J_1$ and $J_2$ which also satisfy 
$$ y^2dx\wedge dy + x^2 dz\wedge dx + z^2 dy \wedge dz = K_1 \wedge K_2 = L_1 \wedge L_2, $$
where $(K_1,K_2) = (y^2 dx - z^2dz, dy - \frac{x^2}{z^2}dx)$ and $(L_1,L_2) = ( z^2 dy - x^2 dx, dz 
- \frac{y^2}{x^2}dy)$ but none of them are closed.

\subsection{Homotopy operator and closed form}

Let $M$ be a manifold and let $I = [0,1]$
Suppose $\omega \in \Omega^k(M \times I)$, every $\omega$ can be uniquely decomposed to
$\omega = \alpha_1 + \alpha_2 \wedge dt$ with $\alpha_1(0,t) \in \Omega^k(M)$ and $\alpha_2(0,t) \in \Omega^{k-1}(M)$.
We define the following mapping
\be \label{hom}
D_k : \Omega^k(M \times I) \to \Omega^{k-1}(M), \qquad (D_k\Omega)(m) : = (-1)^{k-1}\int_{0}^{1}\alpha_2(m,t) dt,
\ee
where the integral is to be understood as an integral of a function on the interval $I$ with values in
the vector space $\wedge^{k-1}T_{m}^{\ast}M$, \cite{RS}. Notice that, this satisfies
\be
d(D_k\omega) + D_{k+1}(d\omega) = \omega|_{t=1} - \omega|_{t=0}.
\ee

\smallskip

Let $\alpha \in \Omega^k(N)$, let $ \phi_{0}, \phi_1 : M \to N$ be smooth mapping,
Then we set $\Omega = F^{\ast}\alpha$ and obtain
\be \label{hom2}
d(D_kF^{\ast}\alpha) + D_{k+1}(dF^{\ast}\alpha) = \phi_{1}^{\ast}\alpha - \phi_{0}^{\ast}\alpha.
\ee
Here, the operator $H_k = D_k \circ F^{\ast}$ is called the homotopy operator.
If $\alpha$ belongs to a cohomology class in $H^k(N)$, then $dD_kF^{\ast}\alpha = \phi_{1}^{\ast}\alpha
- \phi_{0}^{\ast}\alpha$. So $\big(\phi_{1}^{\ast}\alpha - \phi_{0}^{\ast}\alpha\big)$ differ by an exact form,
hence they define the same cohomology class.

\smallskip

The operator $D_k$ defined in (\ref{hom2}) yields an explicit potential form of a closed
form on a contractible manifold. We recall that if $M$ is contractible then by Poincar\'e lemma
$H^k(M) = {\Bbb R}$ for $k =0$ otherwise it is zero for all other $k > 0$.
 Let $F : M \times I \to N$ be a homotopy fulfilling
$F(m,0) = \phi_0(m)$ and $F(m,1) = \phi_1(m)$, where $I =[0,1]$.
From (\ref{hom2}), we obtain
$$ dD_kF^{\ast}\alpha + D_{k+1}dF^{\ast}\alpha = -\alpha, $$
for all $\alpha$, $d\alpha = 0$, then, 
\be
\beta = -D_kF^{\ast}\alpha.
\ee
\begin{claim}
Let 
$\alpha = y^2 dx \wedge dy + x^2 dz \wedge dx + z^2 dy \wedge dz$ be a two form and 
$F((x,y,z),t) = (tx,ty,tz)$ be a homotopy mapping. Then the exact one form $\eta = -D_2F^{\ast}\alpha$
is given by
\be
\eta = - \frac{1}{4}\big( (y^3 - x^2z)dx + (z^3 - xy^2)dy + (x^3 - z^2y)dz \big),
\ee
such that $\alpha = d\eta$. 
\end{claim}
Notice that $\eta= \eta_1 dx + \eta_2 dy + \eta_3 dz$ plays the role of Hamiltonian (one) form. 

Let us consider first order Hamiltonian equations in 2D, we can express it
$$
\dot{x}\Omega = dx \wedge dH, \qquad \dot{p}\Omega = dp\wedge dH, \,\,\,\,\hbox{ where }\,\,\,\,
\Omega = dx \wedge dp,
$$
which yields Hamiltonian equations in the standard form 
$\dot{x} = \frac{\partial H}{\partial p}$ and $\dot{x} = -\frac{\partial H}{\partial x}$.
Similarly, we can express 3D equations of motion as
\be
\dot{x}_i{\tilde\Omega} = dx_i \wedge d\eta, \qquad \hbox{ where } \,\,\,\, {\tilde\Omega} = dx \wedge dy \wedge
dz,
\ee
where $x_i = x,y,z$. Expanding in components we obtain
\be
\dot{x} = -\frac{\partial \eta_2}{\partial z} +  \frac{\partial \eta_3}{\partial y}, \qquad 
\dot{y} = -\frac{\partial \eta_3}{\partial x} +  \frac{\partial \eta_1}{\partial z}, \qquad 
\dot{z} = -\frac{\partial \eta_1}{\partial y} +  \frac{\partial \eta_2}{\partial x}.
\ee
This set of equations are also obtained by Dumachev \cite{Duma1,Duma2} and he called as vector Hamiltonian
equation.

\subsection{Deformation and Hamiltonization}

In this section we deform the two form (\ref{eqntwoform}) in such a way that the dynamics involved
in the modified system also yields velocity as a curl of potential vector field and during this process we will get rid of
null condition, i.e. $\mathbf{A}\cdot \nabla \times \mathbf{A} = 0$. 
Let us assume one form $\tilde{J}_1 = xdx + ydy + zdz$
and demand another form $\tilde{J}_2 = A_1 dx + A_2 dy + A_3 dz$ in such a 
way that it would yield the original form (\ref{eqntwoform}) and
at the same time it should be exact. We arrive at the following set of (deformed) equations
\be\label{deformeqn}
\dot{x} = z^2 - y^2 + xz - xy, \,\,\,\,\, \dot{y} = x^2 - z^2 + xy - yz, \,\,\,\,\, \dot{z} = y^2 - x^2 + yz - xz.
\ee
It is easy to check that (\ref{deformeqn})  yields a divergence free vector field $\mathbf{v}$
and it satisfies $\mathbf{v} = \nabla \times \mathbf{A}$, where the vector potential
$\mathbf{A}$ is given by
\be
\mathbf{A} = \big( (y^2 + z^2)x + xyz, (x^2 + z^2)y + xyz, (x^2 + y^2)z + xyz \big).
\ee  
It is easy to check that the vector potential $\mathbf{A}$ for the deformed equation
satisfies $\mathbf{A}\cdot \nabla \times \mathbf{A} \neq 0$.

\smallskip

\noindent
The identity (\ref{dv}) becomes
\be\label{eqntwoform1}
\big(y^2 - x^2 + yz - xz \big) dx\wedge dy + \big( x^2  - z^2 + xy - yz \big) dz\wedge dx 
+ \big(z^2 - y^2 + xz - xy \big) dy \wedge dz = \hat{J}_1 \wedge \hat{J}_2,
\ee
where $\hat{J}_1$ and $\hat{J}_2$ are exact one forms, $\hat{J}_1 = dI_1$ and $\hat{J}_{2} = dI_2$
with the potential functions
\be\label{Inew}
I_1 = xy + yz + zx, \qquad I_2 = \frac{1}{2}( x^2 + y^2 + z^2).
\ee 
Using these two first integrals, we can express equation (\ref{deformeqn}) in the Nambu-Hamiltonian form (\ref{NH3D}) as follows
\be
\mathbf{\dot{x}} = \{\mathbf{x},I_1,I_2\}=\nabla I_1 \times \nabla I_2
\ee
where $I_1$ and $I_1$ are the ones in (\ref{Inew}).

\section{Outlook}
We studied Hamiltonian aspects of divergence-free vector fields in dimension $3$, chaotic aspects of these kind of 
equations have been studied in \cite{TangBoozer,ThiffeaultBoozer}, in general handful of papers are known 
in the literature for three-dimensional divergence-free vector fields.
In particular, we have studied $\dot{\bf x} = \nabla \times \mathbf{A}$ type flows,
and all these flows satisfy Frobenius integrability condition in a sense that
$i_{\mathbf v}\Omega \wedge J = 0$, or in other words, $i_{\mathbf v}\Omega = J \wedge K$,
where $J$ and $K$ are 1-forms. We explored that not all the flows yield (Nambu) Hamiltonian
framework, it depends on the nature of $J$ and $K$, whether they are closed or not.
We have demonstrated that when we  
deform the second class of system to get rid of
null condition $\mathbf{A} \cdot \nabla \times \mathbf{A} = 0$, the system possesses the Hamiltonian realization. In this paper we could not prove the existence theorem for Hamiltonian
framework but instead of that we have demonstrated the existence of both Hamiltonian and non-Hamiltonian
type flows for $\mathbf{v} = \nabla \times \mathbf{A}$ type 3D flows. Also, very little is known 
about divergence-free vector fields in dimension $n \geq 4$, so we will focus on this problem in our next project.
This piece of work also raised several questions regarding the
applicability of the Euler theorem of potential.

\section*{Acknowledgements}
We would like to express our sincere appreciation to
Professors Sir Michael Berry, Tony Bloch, Larry Bates and Jean-Luc Thiffeault for their
interest and valuable comments. PG is also grateful to 
Vishal Vasan for enlighting discussion. This work has been done while PG is visiting Gebze Technical University, Department of Mathematics,
under TUBITAK 2221 Fellowships for Visiting Scientists and Scientists on Sabbatical Leave program.
He would like to express his sincerest gratitude to all members of department for their warm hospitality,
especially to the chairman Mansur Hoca.

\end{document}